# Making Sense of the Current Covid 19 Situation and Suggesting a tailored Release Strategy through Modeling And Simulation Case Study: Casablanca, Morocco

Mohamed Amine Chadi
*LISI Laboratory, Computer Science Department, FSSM,
Cadi Ayyad University,*
Marrakech, Morocco
m.aminechadi@gmail.com

Hajar Mousannif
*LISI Laboratory, Computer Science Department, FSSM,
Cadi Ayyad University,*
Marrakech, Morocco
hajar.mousannif@gmail.com

*Abstract*— Since the very first days of announcing Covid 19 as a global pandemic, researchers around the world from different backgrounds and organizations including national governments and the World Health Organization (WHO) have extensively begun using Modeling and Simulation (M&S) techniques in deciding on the optimal strategies in order to mitigate the public health and economical effects of Covid-19. In this paper, we present our results from applying M&S on the city of Casablanca, Morocco to comprehend (i): How it can be relatively easy or difficult for such pandemic to penetrate to the studied region (country, city, district, etc.). (ii): How it can spread relatively fast or slow within that region. And finally (iii): What is the optimal strategy for containing and mitigating the propagation of the disease while still keeping the economy as near to normal as possible.

*Keywords— Covid-19; Modeling and simulation; Lockdown-release strategy; Morocco, Casablanca*

## I. Introduction

COVID-19 is certainly one of the worst pandemics ever. As of 20 April 2020, the virus has generated over 2 million of confirmed infected cases and 150 thousand deaths [1]. In the absence of a vaccine, classical epidemiological measures such as testing in order to isolate the infected people, quarantine and social distancing are ways to reduce the growing speed of new infections as much as possible and as soon as possible. But, the economical harm of these measurements is estimated to be much more than that of the economical crisis of 2008 [2]. On the other hand, some studies [3] suggest that quarantine and social distancing measures might be needed for as long as 18 months, and will only be turned on and off alternatively during that period.

In order to get a better understanding and make quick and optimal reactions to the current situation, many researchers have decided to exploit the various M&S tools and techniques [4][5][6].

In this paper, we aim to explain through M&S the current situation of the Covid 19 pandemic in Casablanca, Morocco, in order to reinforce the "so far" efficient measures, and also suggest a context tailored optimal strategy of release in order to contain and mitigate the spread of the virus while keeping the economy as near to normal as possible.

The urgency of the current situation and the fact that the targeted reader is meant to be from any background related to this subject is what guided the plan of this article to be as straight-forward, concise and practical as it can be. Therefore, section 2 is a brief presentation of the studied region (the city of Casablanca), its eight districts and some of their socio-economical characteristics. Section 3 discusses the current situation of Covid 19 in Casablanca. Section 4 is a walkthrough the simulations techniques used to explain the previously presented pandemic's situation. And finally in section 5 we suggest (what we believe to be) an optimal strategy for the after lockdown time as the Moroccan government is planning for a release in 20 May 2020.

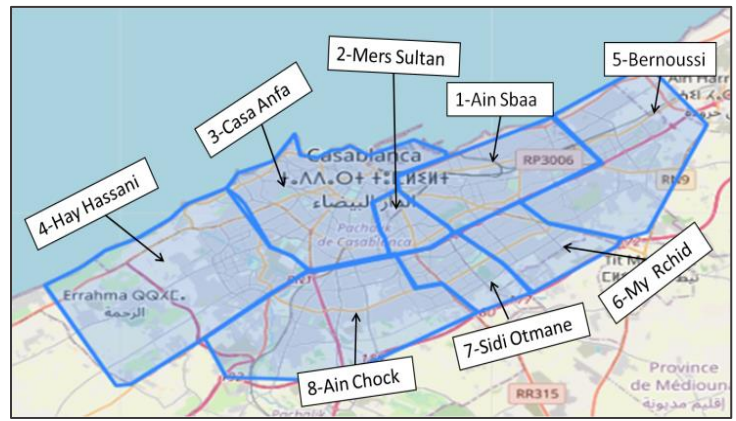

**Figure 1: Map of Casablanca city and its eight districts**

## II. About The City Of Casablanca

Casablanca is the metropolitan economical capital of Morocco, most densely populated and currently, it holds the highest number of infections. Casablanca has eight districts (see figure 1), each district in Casablanca has its own demographic characteristics such as population size, density and dynamics but they don't differ that much. What might be the main difference between them is the type of each district (industrial, downtown, etc.)). But to summarize all briefly, there are three categories:

1st category: **Casa Anfa** district serves as the city's economic and administrative center, cultural identifier, and historical birthplace.

2nd category: The three districts **Ain Sbaa**, **Ain Chock** and **Bernoussi** are host to lots of industrial areas and businesses.

3rd category: the remaining four districts: **My Rchid, Sidi Otmane, Mers Sultan** and **Hay Hassani** are mostly popular middle and working-class districts.

Each category of districts maps into one of three classes, this classification is based on the number of transportations quantity and means provided for a certain district and also on the number of reasons people would go there, such as super market, work, hospitals, etc. which can be found in the official website of the administrative division of Casablanca here [7].

-**Highly attractive**: that is a district that people from all other districts go to for lots of reasons, in this category, only *Casa Anfa* district is put here.

- **Moderately attractive**: these are districts that present a relatively smaller number of reasons to visit; mostly in our case is the number of companies existing, this category holds *Ain Chock, Bernoussi* and *Ain Sbaa* districts.

- **Slightly attractive**: this category is for the last four districts *My Rchid, Sidi Otmane, Mers Sultan* and *Hay Hassani*.

### III. THE CURRENT SITUATION OF COVID 19 IN CASABLANCA

The table below shows the number of the infected people in each district as of 19 April 2020.

Even though the earlier cases were reported in Sidi Otmane and My Rchid districts, after 6 weeks now, both of the two districts have very low number of cases (See table.1).

**Table 1: the cumulated number of confirmed cases per district as of 19 April 2020**

| District | Number of cases |
|---|---|
| Casa Anfa | 122 |
| Ain Sbaa | 102 |
| Ain Chock | 87 |
| Bernousi | 77 |
| My Rchid | 44 |
| Hay Hasani | 41 |
| Mers Sultan | 40 |
| Sidi otmane | 30 |

One thing to mention, this is not because people from these two districts respected the lockdown measurements the most. In fact the number of reported violations of the measurements is the highest among those two districts more than others. So why this is the case!

Given that all districts have quite similar demographics, everyone expect that the one that starts first should be at the top of the ranking list.

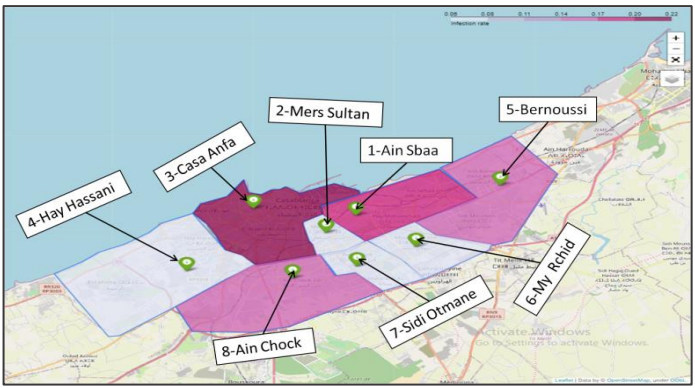

**Figure 2: the choropleth map of the distribution of the infection rate in each district of Casablanca**

And here is the choropleth map version of the above table:

The choropleth map (figure 2) is utilized in order to easily represent the variation of the infection rate across the eight districts of the city of Casablanca. It is actually what made us question the hidden reasons why some districts are much more impacted than others even if they are near to each other. For example, Mers Sultan district happened to be located right between the two most infected districts among all, yet it is surprisingly at the bottom of the ranking table.

### IV. MAKING SENSE OF THE DATA

*1) First simulation:*

Surely, by now everyone interested in modeling and simulation of Covid 19 has seen the Washington Post article by Harry Stevens [8]. Inspired by it, we considered each district in our previous map (figure 1) as a box containing 100 agents bouncing around randomly with 50% immobility. If there is one thing we should learn from this simulation, is that the number of infected people next day depends mainly on: (i) the initial number of infectious people and (ii) the contact rate (i.e. density and movement rate).

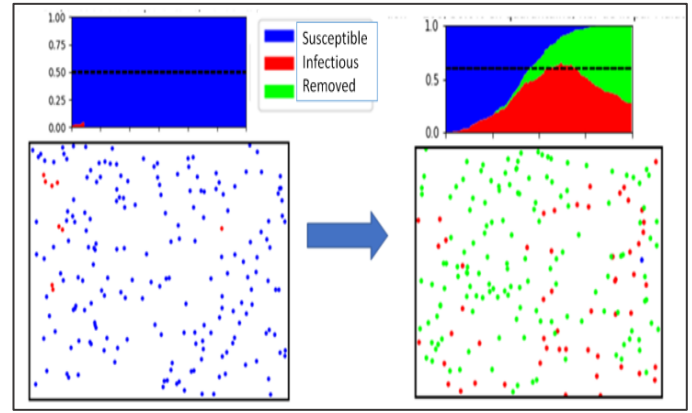

**Figure 3: the propagation of an epidemic proportional to the contact rate and initial number of infected people**

*2) Second simulation:*

**a) Definition:**

*Connection rate:* is essentially how much important is a district to people from other districts, this can be estimated from how much vital sectors (hospitals, companies, businesses, etc.) are located within this district, adding transportation's destination data, which both are available at [7] the official website of the administrative division of Casablanca.

**b) Simulation:**

Here, we consider that all districts have same number of population (100) and same number of initial infectious people (3).

The box with the largest border line is the most connected/attractive district (In our case, it is Casa Anfa). Then there are three districts with thinner border line, those are work areas (Ain Sbaa, Ain Chock and Bernoussi). The other four boxes are the remaining districts that are considered to have very low connection rate.

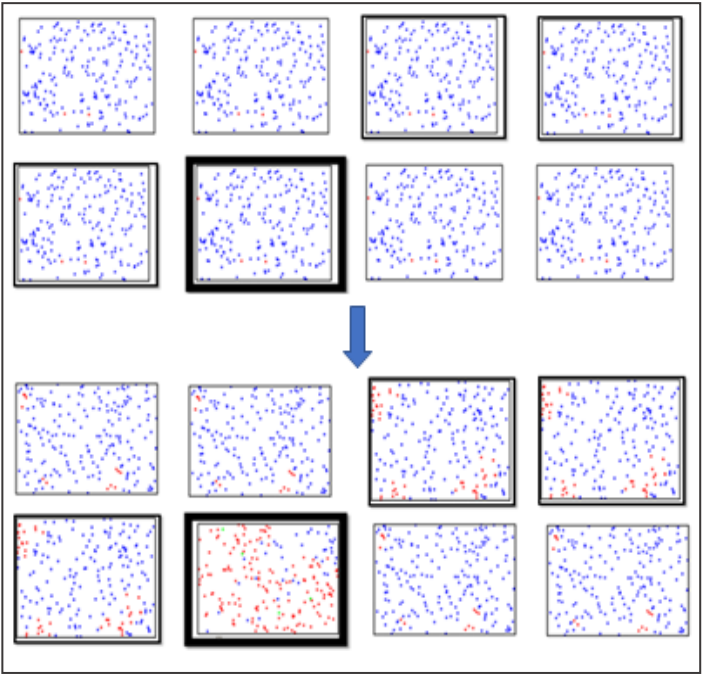

**Figure 4: evolution of an epidemic in city's center, work areas and popular areas**

Because the city's center receives the 3 infected from each other region, city's center is now with 3x7 = 27 infected. Being as dynamic as such places, the contact rate tend to be much higher, inducing a high infection rate per period. Now Casa Anfa should be shut down. Consequently, economy will decline, thus no budget to buy health equipments and eventually the mortality rate will get very high. And same logic goes for the work places, it just would be a little lower because not everyone works in the same place but instead the effect is distributed among all three districts.

Also, the contact rate is relatively low in the four popular districts of the third category. This is due to the fact that people that are living in these areas and have regular jobs, their job's place is most likely located either in the city's center or in the three districts with work areas.
Therefore, a considerable number of people spend most of the daytime in other four districts and not in their districts, and only after 6PM that they return. And because of lockdown measures, after 6PM no one is allowed to stay out, thus, the coming back people goes right to their homes after spending the whole daytime in their work places.

## V. DISCUSSING THE OPTIMAL LOCKDOWN-RELEASE STRAREGY

*1) Context:*
While some research [9] argues that a young-workforce release would lead to substantial economic and societal benefits without enormous health costs to the country, others [10] suggest a community-based incremental approach to the easing of lockdowns, tailored to demographic and social stratifications of risk.

Here, we present a more flexible mix of the above propositions with slight changes and added tips in order to fit the context of a developing country like Morocco.

*2) The lockdown-release strategy:*
**1-A new wave of infections after Release from lockdowns is the "most probable" scenario!**

Unless, either the virus ran out of people to infect or a vaccine is available, no matter how much people respects social distancing measures, at the end, if only one person was infected and had not been identified, the wave of infections will start again. Many M&S driven studies [11][12][13] predict that any non careful release after lockdowns will result in a new wave of infections

**2- How to prevent that:**

As presented by the 2018 Nobel Prize laureate, economist Paul Romer [14], testing and isolating everyone every short period of time (each week or even each day) is certainly the best straight-forward strategy to prevent the society from a new wave of infections. But given the fact that it's still a debate whether it is feasible or not in terms of the cost needed to make it in the United States, it is certainly not feasible to make it with that scale in a developing country such as Morocco. Thus, in the following, we present an **adapted hybrid incremental strategy for lockdown-release** for the Moroccan-like countries.

**a) Identify highly attractive areas (with highest connection rate):**

In the majority of cases, these are areas where big companies, businesses and commercial centers are located, which means that they present a high contact rate. In addition, these areas are critical to the maintenance of the global economy, therefore identifying them can certainly be of huge benefit in mitigating the spread of the virus as well as maintaining the economy.

In our case, the area that presented the highest connection rate was Casa Anfa. This is because of the number of businesses, companies, government and private hospitals, high-quality education private schools, etc. This assumption was later verified by analyzing the transportation data in the official website of the administrative division of Casablanca [7]. The majority of transportation means from all other districts are those that have a destination to Casa Anfa (figure 4).

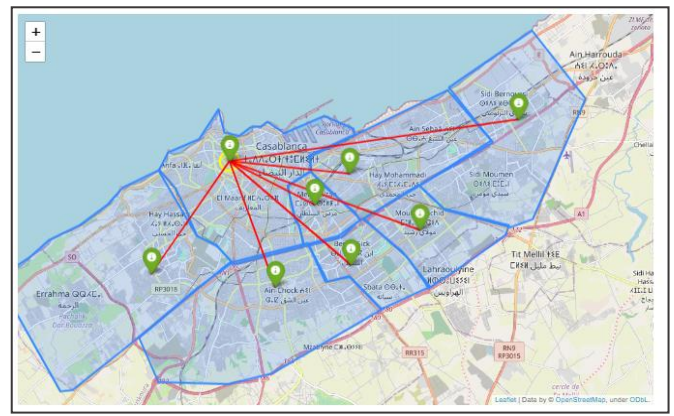

**Figure 5: Map to network for the most attractive district (Casa Anfa)**

**b) Test, Isolate and trace within the identified areas:**

Until the Moroccan authorities have a sufficient amount of testing and isolation hotels capacity alongside with a contact tracing system for a targeted and non wasted test; **not for the whole population**; but only for people who live and/or work in the previously identified essential areas for the global economy, we believe that Morocco cannot reopen without having a new wave of infections as the most probable scenario.

Once we are pretty sure the identified areas are totally virus-free, we should not allow any visitors to come-in unless they are confirmedly tested negative for the maintenance of the public health of these areas, as well as the global economy.

**c) Generalize for other cities of the country gradually:**

Implementing this strategy in other metropolitan cities of the country an incremental manner is highly recommended since it will much more strengthen the global economy until it is near normal even with this infectious disease hanging around.

Another consideration might be to release the youth workers (20-30 years old) that are living apart from vulnerable people such as their parents, especially if they represent a significant proportion.

**d) Now that the economy is "near" normal, what's next?**

At this point in time, we shall begin releasing gradually people from the other areas while still keeping some of the essential good practices and measurements such as wearing masks and hand washing. Even if someone was tested positive now, we know for sure that this infected person did not go to the identified economical areas while he was infected (because we test anyone coming in).

In addition, even in the worst-case scenario, if an infected person transmitted the disease to lots of people, we now would have a good economy to assess the health system.

**e) Argument about the budget:**

Surely, this strategy would require a big budget at the beginning for buying a large amount of testing kits (or making them locally), but let's think about it for a second, it is far much better to spend 1 Billion "for once", than to keep spending 400 Million each time for multiple lockdowns and releases loop.

## VI. CONCLUSION AND FUTURE WORK:

In this paper, through a M&S driven analysis of the existing data of Casablanca city epidemic situation, we studied the main factors affecting the spread of COVID-19, leading us to determine three parameters as the most important ones: the number of infected people now, the contact rate and connection rate.

What's more, is that our models helped us in inferring the reasons why some regions might be more susceptible for easy penetration and rapid spread of the virus than other regions. Not only that, but it also helped us to suggest what might be the optimal policy in order to contain and mitigate the spread of the virus while still maintaining the global economy as near to normal as possible.

Finally, we hope that this article can make some contributions to the world's response in general and Moroccan's response in particular to this epidemic and give some references for future research.

Future research will focus on applying the same approach described here for the prediction of the epidemic's evolution in other Moroccan and international regions (mainly cities) as well as adding other features related to optimal lockdown release strategies.

### Important note :

This manuscript reports an early version of the model and results, while deeper analysis is in progress.

We wish to disseminate this work as soon as possible, in the context of the current Covid-19 urgency in general and for our country of Morocco in particulier. We will continue to add more detailed description and literature review in stages.

An interactive version of the simulations used in this paper is systematically published in :
https://covid.mousannifhajar.com/

And the source code is available on github at :
https://github.com/amine179/Covid19_Morocco